\documentclass[twocolumn]{aastex63}
\usepackage{amssymb,amsmath}
\usepackage{graphicx}
\usepackage{xcolor,natbib}
\usepackage{multirow}
\usepackage[T1]{fontenc}
\usepackage{ae,aecompl}
\usepackage{newtxtext, newtxmath}
\usepackage{float}
\usepackage{subfigure}
\usepackage{longtable}
\usepackage{enumitem}
\usepackage{longtable,listings}
\usepackage[flushleft]{threeparttable}
\usepackage{parcolumns}
\usepackage{makecell}

\def\nii{[N~{\sc ii}]$\lambda6583$\AA}
\def\si{[S~{\sc ii}]$\lambda6716$\AA}
\def\sii{[S~{\sc ii}]$\lambda6732$\AA}
\def\oc3{[O~{\sc iii}]}
\def\o4{[O~{\sc iii}]$\lambda4364$\AA}
\def\oe3{[O~{\sc iii}]$_{ext}$}
\def\oi{[O~{\sc i}]$\lambda6300$\AA}
\def\oii{[O~{\sc i}]$\lambda6363$\AA}
\def\obj{SDSS J1346+1736}

\submitjournal{ApJ}

\shorttitle{moving dust clouds for CLAGN}
\shortauthors{Zhang XueGuang}

\begin{document}

\title{Unique photometric variability in SDSS J134628.62+173659.5: clues for moving dust clouds as physical origin of 
changing-look AGN}

\correspondingauthor{XueGuang Zhang}% \email{xgzhang@njnu.edu.cn}}
\email{xgzhang@gxu.edu.cn}
\author{XueGuang Zhang$^{*}$}
\affiliation{Guangxi Key Laboratory for Relativistic Astrophysics, School of Physical Science and Technology, GuangXi
University, Nanning, 530004, P. R. China}

 %%%about 154 words
\begin{abstract}
The scenario of variations in accreting process around central black hole has been widely accepted as the preferred physical 
origin of changing-look active galactic nuclei (CLAGN), rather than obscuration effects by moving dust clouds. In this manuscript, 
after analyzing long-term photometric variability in Type-1.8 AGN SDSS J1346+1736 with apparent broad H$\alpha$ 
but very weak broad H$\beta$, robust clues for obscurations can be confirmed with confidence level higher than 10$\sigma$. Then, 
based on obscurations related to moving dust clouds, from dark state with E(B-V)$\sim$1.26 to bright state with E(B-V)$\sim$0.55, 
apparent both broad H$\alpha$ and broad H$\beta$ can be clearly expected. Meanwhile, the expected line luminosity from reddening 
corrected continuum luminosity is consistent with the reddening corrected line luminosity, to support the obscuration effects. 
Furthermore, through symmetric features in the light curves, probability is only 3.2\% in SDSS J1346+1736 to support the scenario 
of variations of accretion rates traced by CAR process. Therefore, besides the more and more preferred scenario of variations in 
accreting process, the scenario of moving dust clouds could be more reasonable in some CLAGN with apparent variations in optical 
colors. 
\end{abstract}


\section{Introduction}

	Changing-look active galactic nuclei (CLAGN) has been commonly accepted as a unique subclass of AGN, due to its optical 
spectral type being changed between Type-1 AGN with apparent broad Balmer emission lines and Type-2/1.9 AGN without apparent broad 
H$\beta$. Since the CLAGN firstly reported around 1980s in \citet{to76, ch86}, through properties of spectroscopic and/or photometric 
variability, there are so far hundreds of CLAGN reported in the literature, such as the individual CLAGN in \citet{sb93, eh01, sp14, 
lc15, run16, wx18, gm19, ta19, zh21, lm22, zl24, ln25, lw25, sm25}, etc., and the samples of CLAGN reported in \citet{mr16, yw18, fg19, 
mg19, sw20, gp22, tr23, aw24, gz24, zt24, lu25}, etc.

	In order to explain the unique spectroscopic variability of CLAGN, the following three scenarios have been mainly considered. 
The moving dust clouds have been proposed and discussed in \citet{gr90, rs09}. The common variations of accretion rates have been 
proposed and discussed in \citet{lw22, yw23, de25, sl25}. And the variations of accretion rates related to tidal disruption events 
have been proposed and discussed in \citet{md15, zh21b, zh22, zh23, wl24, zh25}. The more recent systematic discussions on CLAGN can 
be found in \citet{zt22, rt23}. With more and more CLAGN detected, the scenario of variations in accretion rates related to central 
intrinsic AGN accreting process or to tidal disruption events has been more and more popular for the physical origin of spectroscopic 
variability properties of CLAGN, rather than the scenario of moving dust clouds, especially after considering the scenario of moving 
dust clouds expected timescales very longer than the observed variability timescales in some CLAGN as discussed in \citet{zt22}.

	Actually, due to few effects of obscurations related to intrinsic AGN accretion variations or tidal disruption events, such 
as the results shown in Fig.~2 in \citet{yao23}, CLAGN if having large variations in optical colors could be applied to support 
central obscurations. Unfortunately, there is so far no one CLAGN clearly reported and discussed with apparent and large variations 
in optical colors, therefore, the scenario of moving dust clouds has been considered less and less. However, when checking gri-band 
light curves from Zwicky Transient Facility (ZTF) \citep{be19, ma19} of low redshift SDSS quasars, the AGN SDSS J134628.62+173659.5 
(=SDSS J1346+1736) at $z\sim$0.17 shows apparent and large variations in its optical colors, indicating serious central obscurations. 
Further analysis of both spectroscopic and photometric properties could provide clues to support the scenario of moving dust clouds 
in CLAGN, which is the main objective of this manuscript.

\begin{figure}
\centering\includegraphics[width = 9cm,height=9cm]{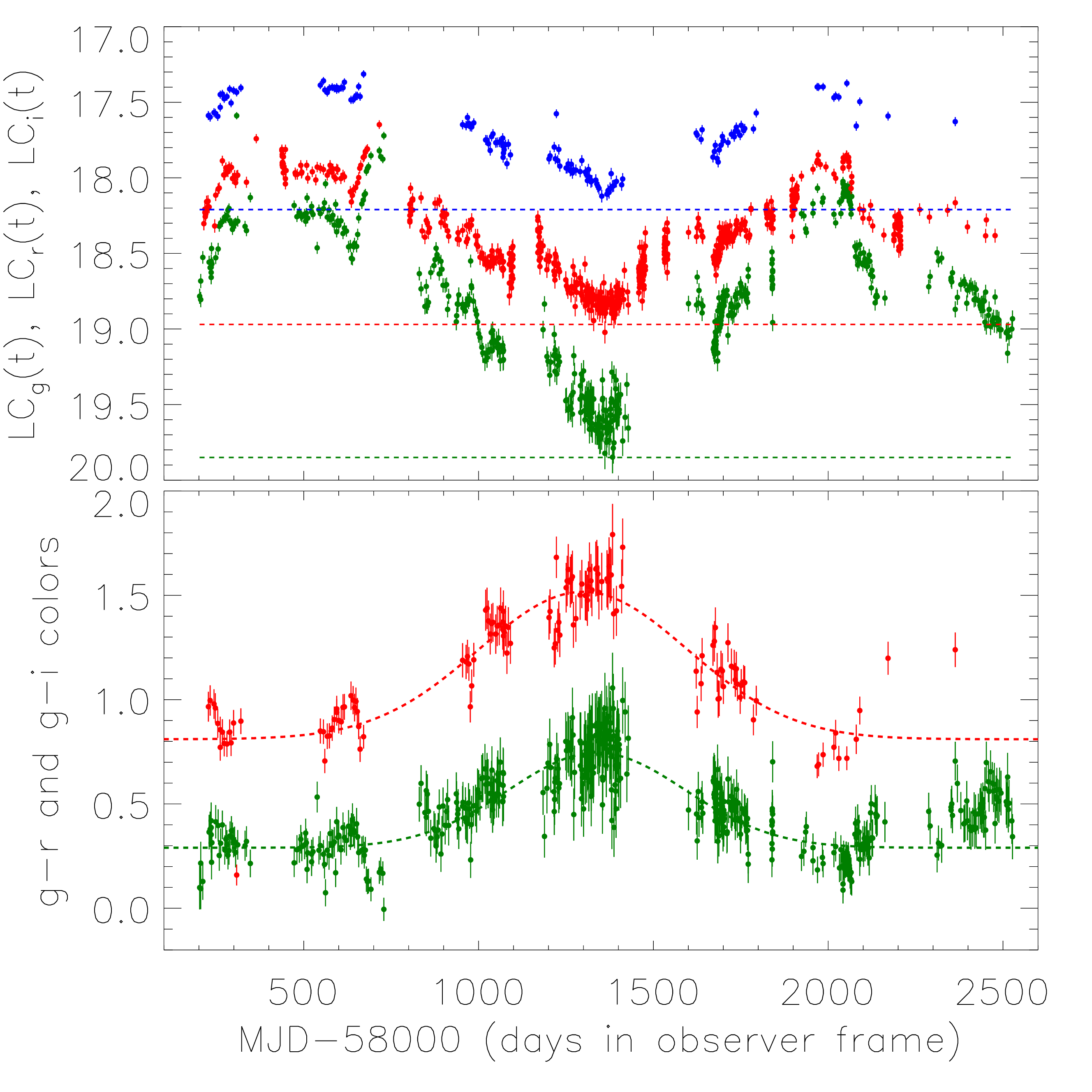}
\caption{Light curves (top panel) and colors (bottom panel) in \obj. In top panel, solid circles plus error bars in dark green, 
in red and in blue show the light curves in ZTF gri-band, respectively. The horizontal dashed lines in dark green, in red and in 
blue mark corresponding ZTF gri-band magnitudes determined through the SDSS spectrum. In bottom panel, solid circles plus error 
bars in dark green and in red show the g-r and g-i colors, respectively, with dashed line in the same color showing the Gaussian-like 
descriptions to the optical color. }
\label{lmc}
\end{figure}

\begin{figure*}
\centering\includegraphics[width = 18cm,height=12cm]{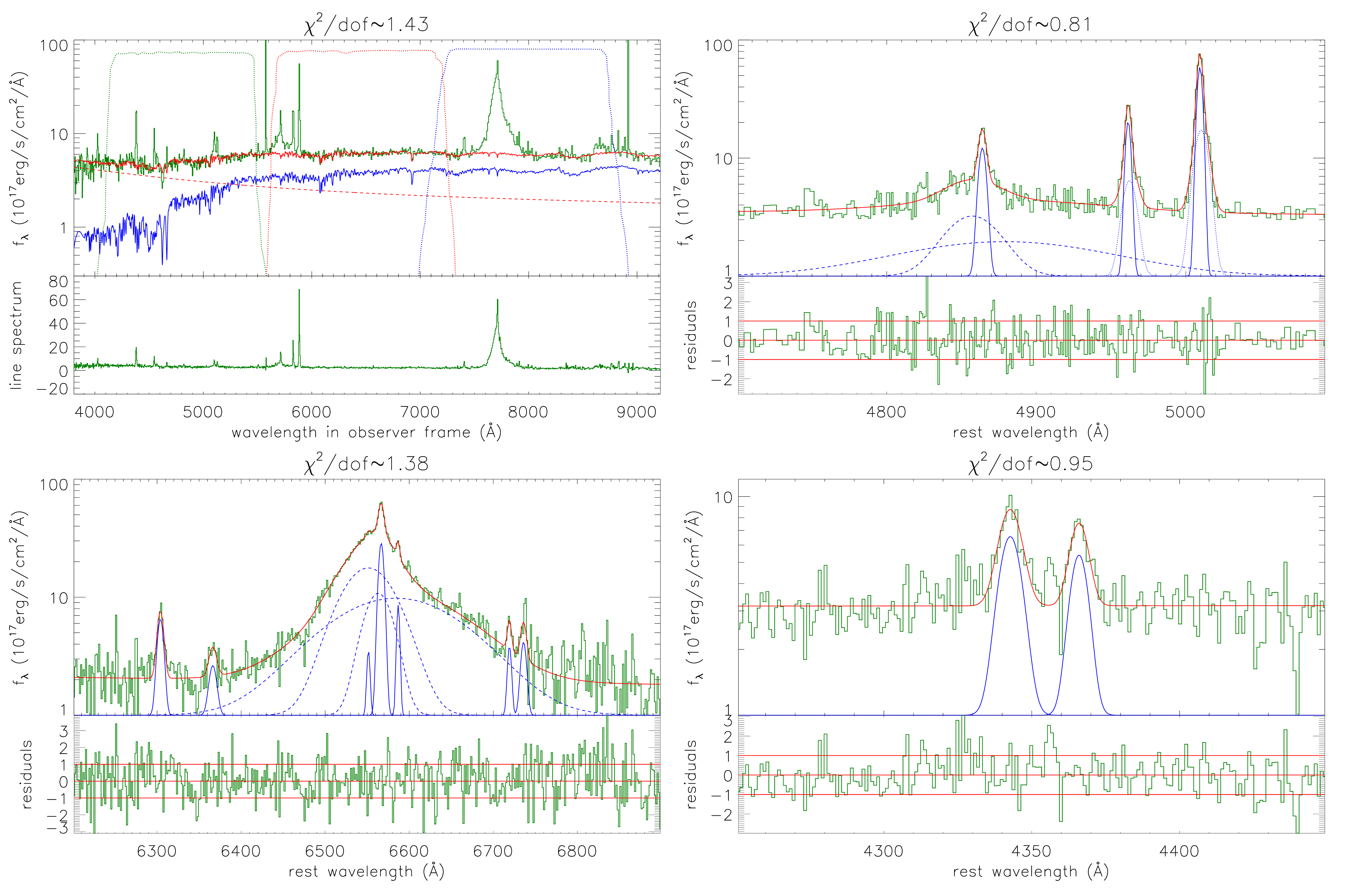}
\caption{Spectroscopic properties of \obj. Top left panel shows the SDSS spectrum (solid dark green line) and the best fitting 
results (solid red line). Solid blue line and dashed red line show the determined host galaxy contributions and AGN continuum 
emissions. Dotted lines in dark green, in red and in blue show the 80times scaled ZTF gri-filter transmission curves. Bottom region  
of the top left panel shows the line spectrum calculated by the SDSS spectrum minus the host galaxy contributions. The other panels 
show the best fitting results (solid red line) to the line spectrum (solid dark green) around H$\beta$, H$\alpha$, H$\gamma$ and 
corresponding residuals. In top right and bottom left panels, dashed blue lines show the determined Gaussian components in broad 
Balmer lines, solid blue lines show the determined Gaussian components in narrow emission lines. In bottom right panel, solid blue 
lines show the determined Gaussian components in narrow H$\gamma$ and [O~{\sc iii}]$\lambda4364$\AA. In top right panel, dotted blue 
lines show the determined extended components in [O~{\sc iii}] doublet. In bottom regions in the top right panel and the bottom 
panels, horizontal red lines show residuals=$0,\pm1$. }
\label{spec}
\end{figure*}

\begin{figure}
\centering\includegraphics[width = 9cm,height=6cm]{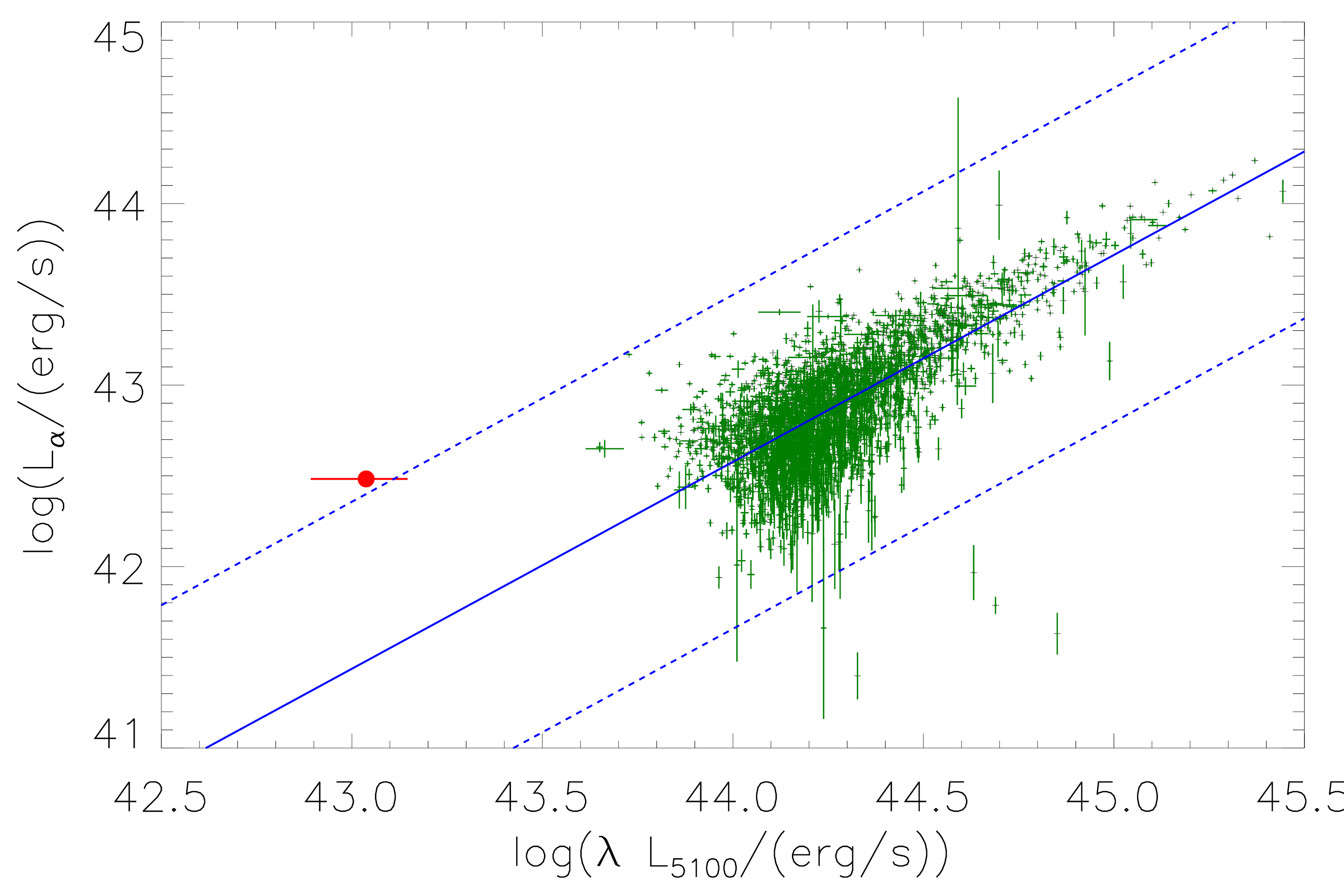}
\caption{On the correlation between continuum luminosity at 5100\AA~ $\lambda L_{5100}$ and the line luminosity $L_\alpha$ (including 
contributions of narrow and broad H$\alpha$ and [N~{\sc ii}]). Symbols in dark green show the results of the selected 3206 SDSS low 
redshift quasars, solid circle plus error bars in red show the results of \obj. Solid and dashed lines in blue show the best 
descriptions and corresponding 5RMS scatters to the correlation, similar as the results reported in \citet{gh05}.}
\label{cl}
\end{figure}

\begin{figure}
\centering\includegraphics[width = 9cm,height=7cm]{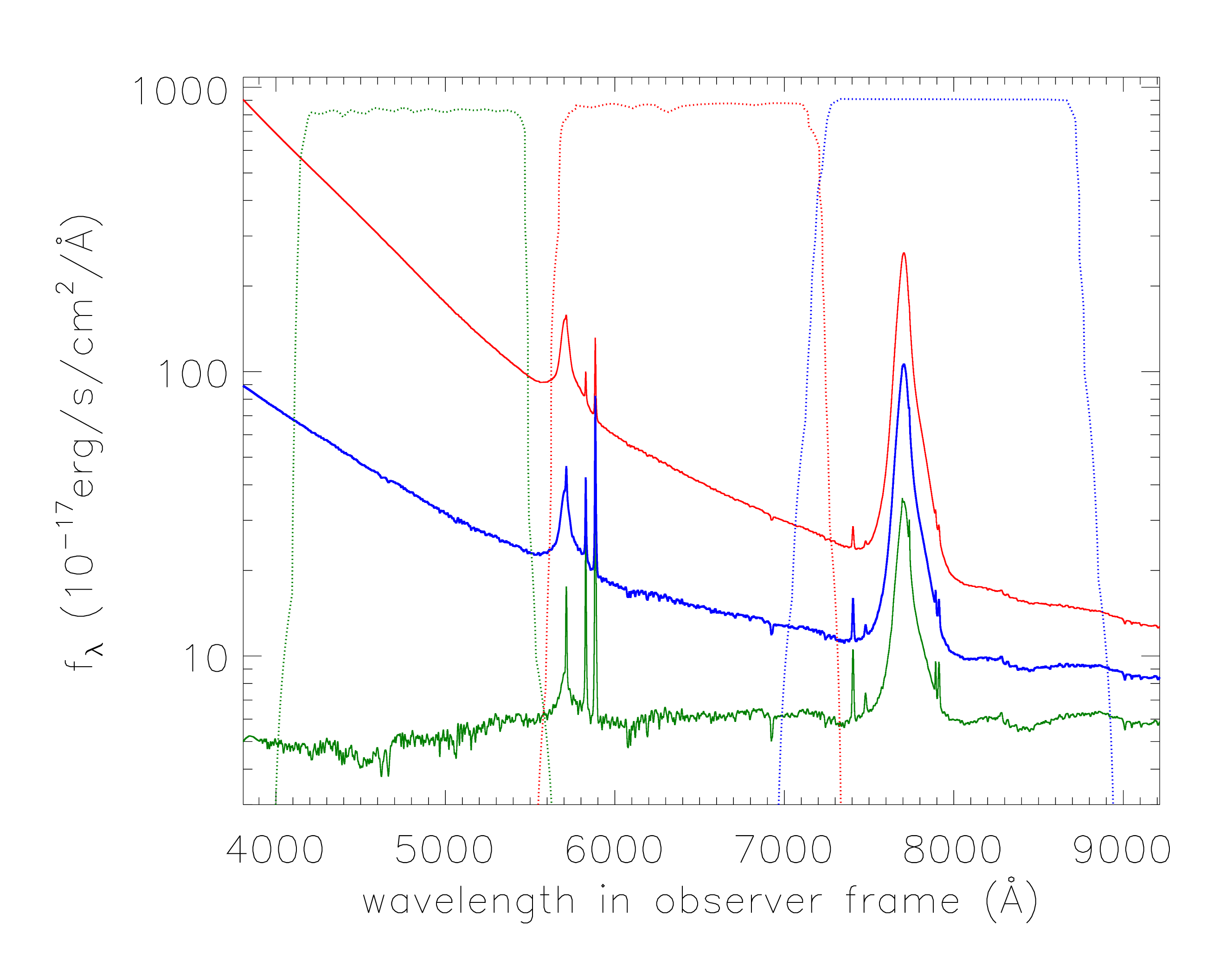}
\caption{Effects of obscurations on spectroscopic properties of \obj. Solid line in dark green shows the origin spectrum from the 
SDSS spectrum, solid lines in red and in blue show the expected spectrum with E(B-V)=1.26, and with E(B-V)=0.55, respectively, to 
show type being changed from Type-1.8 to Type-1. Dotted lines in dark green, in red and in blue show the 905 times 
scaled ZTF gri-filter transmission curves.}
\label{ebv}
\end{figure}

        This manuscript is organized as follows. Section 2 presents the main results and necessary discussions on both spectroscopic 
and photometric properties of \obj. Section 3 gives the main conclusions. And throughout this manuscript, the cosmological parameters 
have been adopted to be $H_{0}=70{\rm km\cdot s}^{-1}{\rm Mpc}^{-1}$, $\Omega_{\Lambda}=0.7$ and $\Omega_{\rm m}=0.3$.

\section{Main Results and necessary Discussions} 

%%%1

	The 6.26 years-long (MJD-58000 from 202 to 2527, from Mar. 2018 to Sep. 2024) optical ZTF gri-band light curves of \obj~ 
are shown in top panel of Fig.~\ref{lmc}. The corresponding g-r and g-i colors are shown in the bottom panel of Fig.~\ref{lmc} with 
large amplitudes around MJD-58000$\sim$1500, indicating serious obscurations in \obj. Meanwhile, based on a linear trend plus a 
Gaussian component, the colors can be described with $\chi^2/dof$ to be 945/533 and 432/135 in g-r and g-i colors, as shown dashed 
lines in bottom panel of Fig.~\ref{lmc}. If only considering a linear trend to describe the colors, the determined $\chi^2/dof$ are 
2123/536 and 1609/138 in the g-r and g-i colors. Through the F-test technique, similar as what we have recently done in \citet{zh25b}, 
the Gaussian component to support apparent variations in colors is preferred with confidence level higher than 10$\sigma$ in \obj. 
Here, the Gaussian descriptions to the optical colors are only applied to confirm the large enough variations, have no any further 
intentions for physical origin of the large variations. Now it is interesting to check effects of obscurations on broad emission 
lines in \obj.

%%%2
	The high quality SDSS spectrum (PLATE-MJD-FIBERID = 2742-54233-0414) of \obj~ is shown in top left panel of Fig.~\ref{spec}. 
Through the SDSS spectrum convolved with the transmission curves of ZTF gri-filters, the calculated gri-band magnitudes are 19.85, 
18.97 and 18.21. Compared with the ZTF gri-band light curves, the SDSS spectrum with MJD=54233 was in dark state. Then, we can check 
properties of the broad emission lines from the dark state to the bright state. 

%%%3
	Before measuring emission lines, the commonly accepted Simple Stellar Population (SSP) method is applied to determine the 
host galaxy contributions. Similar as what we have recently done in \citet{zh24}, the broadened, strengthened and shifted 39 stellar 
templates \citep{bc03, ka03} plus a 5th-order polynomial component are applied to describe the host galaxy contributions and the 
AGN continuum emissions in the SDSS spectrum with emission lines being masked out. The best fitting results and the corresponding 
line spectrum (SDSS spectrum minus the determined host galaxy contributions) are determined and shown in the top left panel of 
Fig.~\ref{spec}, through the Levenberg-Marquardt least-squares minimization technique. And the measured stellar velocity dispersion 
(the broadened velocity of the templates) is 90$\pm$29km/s, leading to the estimated black hole mass 
$5.5_{-4.4}^{+12.1}\times10^6{\rm M_\odot}$ through the M-sigma relation in \citet{kh13, bt21, zh24b}.

%%%4
	After subtracting the host galaxy contributions, emission lines with rest wavelength from 4600\AA~ to 5300\AA~ (around 
H$\beta$) and from 6200\AA~ to 6900\AA~ (around H$\alpha$) and from 4250\AA~ to 4450\AA~ (around H$\gamma$) are measured by multiple 
Gaussian functions as follows. For each broad Balmer line, three Gaussian functions are applied. For each narrow emission line, one 
narrow Gaussian function is applied, besides additional extended components applied in [O~{\sc iii}] doublet. When the Gaussian 
functions are applied, only two criteria are applied. First, each Gaussian component has emission intensity not smaller than zero. 
Second, the narrow components in [O~{\sc iii}] ([N~{\sc ii}]) doublet have flux ratio fixed to be 3. Then, through the 
Levenberg-Marquardt least-squares minimization technique, the best fitting results and the corresponding residuals are determined 
and shown in top right panel and bottom panels of Fig.~\ref{spec}. Here, the residuals are calculated by the line spectrum minus 
the best fitting results and then divided by uncertainties of the line spectrum. The determined parameters and corresponding 
1$\sigma$ uncertainties of each Gaussian component are listed in Table~1. Here, only two components in broad H$\beta$ can be 
determined with reliable parameters, and there are no broad components determined with reliable parameters in broad H$\gamma$. Here, 
the objective to measure line properties of H$\gamma$ is to classify the \obj through line properties of higher-order Balmer lines 
as described in \citet{rc16}. Meanwhile, due to broad but weak H$\beta$, there are large uncertainties of corresponding fluxes of 
the two broad components of H$\beta$. 

%If considering measured flux smaller than 3 times of their uncertainties in the two broad components in H$\beta$

\begin{table}
\caption{Measured line parameters of emission components}
\begin{tabular}{llll}
\hline
line & $\lambda_0$ & $\sigma$ & flux  \\
\hline
\multicolumn{4}{c}{broad emission components} \\
	H$\beta_{B1}$ & 4857.2$\pm$2.4   &    17.4$\pm$3.1    &   97$\pm$25 \\
	H$\beta_{B2}$ & 4879.1$\pm$16.7   &    66.4$\pm$23.9     &  159$\pm$64 \\
	H$\alpha_{B1}$ & 6551.3$\pm$1.7  &    30.4$\pm$1.3    &    1271$\pm$132 \\
	H$\alpha_{B2}$ & 6563.1$\pm$1.3  &    15.9$\pm$1.9    &    390$\pm$111 \\
	H$\alpha_{B3}$ & 6584.4$\pm$2.4  &    74.2$\pm$2.2    &    1638$\pm$65 \\
\hline
\multicolumn{4}{c}{narrow emission  components} \\	
	H$\gamma_{N}$ & 4342.7$\pm$0.3 & 3.6$\pm$0.3 & 51$\pm$4 \\
	\o4 & 4365.9$\pm$0.4 & 2.9$\pm$0.3 & 32$\pm$3 \\
	H$\beta_{N}$ & 4863.7$\pm$0.2 &      2.1$\pm$0.3   &    69$\pm$5 \\
	\oc3 & 5009.4$\pm$0.1  &     1.8$\pm$0.1  &     258$\pm$21 \\
	\oe3 & 5010.3$\pm$0.2  & 4.4$\pm$0.4  &      179$\pm$21 \\ 
	\nii & 6586.8$\pm$0.2  &     1.7$\pm$0.2  &     33$\pm$4 \\
	H$\alpha_N$ & 6566.6$\pm$0.1  &     3.1$\pm$0.1  &     216$\pm$12 \\
	\si & 6719.2$\pm$0.6   &    2.2$\pm$0.6   &    16$\pm$4 \\
	\sii & 6736.1$\pm$0.7  &     3.2$\pm$0.6  &     25$\pm$6 \\
	\oi & 6303.9$\pm$0.3   &    3.5$\pm$0.3   &    49$\pm$4 \\
	\oii & 6366.3$\pm$1.1  &     4.2$\pm$1.0  &     17$\pm$4 \\
\hline
\end{tabular}\\
Note:The first column shows which component is measured. The suffix 'B' and 'N' mean broad and narrow components in Balmer lines. 
\oc3 and \oe3~ mean the core component and the extended component in [O~{\sc iii}]$\lambda5007$\AA. The second, the third and the 
fourth column show the measured line parameters of central wavelength in units of \AA, line width (second moment) in units of \AA, 
and emission flux in units of $10^{-17}{\rm erg/s/cm^2}$.
\end{table}

%%%5
	Based on the measured broad Balmer lines, the flux ratio (Balmer decrement) is $BD=12.9_{-4.2}^{+8.71}$ of broad H$\alpha$ 
to broad H$\beta$, after considering the uncertainties of emission fluxes of the broad components. The large $BD$ is consistent with 
the expected results of \obj~ having its spectrum observed in the dark state with serious obscurations, as a Type-1.8 AGN 
after considering weak broad H$\beta$ but no broad components in higher-order Balmer lines in \obj~ as described in \citet{rc16}. 
Based on the measured $BD$, corresponding $E(B-V)\sim1.26_{-0.34}^{+0.44}$ can be determined, if accepted the intrinsic value 3.1 
for the flux ratio of broad H$\alpha$ to broad H$\beta$. Meanwhile, based on the measurements of the narrow Balmer lines, the flux 
ratio is $3.13_{-0.37}^{+0.43}$ of narrow H$\alpha$ to narrow H$\beta$, consistent with the Case B expected value. Therefore, there 
are serious obscurations on broad emission lines from central broad emission line regions (BLRs) and on AGN continuum emissions, 
but no obscurations on narrow emission lines or on host galaxy of \obj.

%%%6
	Before proceeding further, two points can be applied to further support the central serious obscurations in \obj. First, 
based on the strong linear correlation between line luminosity and continuum luminosity in \citet{gh05} for unobscured quasars, the 
expected total line luminosity is $4.04_{-1.66}^{+1.36}\times10^{41}{\rm erg/s}$ in \obj~ with measured continuum luminosity 
$(1.09\pm0.31)\times10^{43}{\rm erg/s}$ at 5100\AA~ in rest frame. Here, the total line luminosity includes contributions of broad 
and narrow H$\alpha$ and [N~{\sc ii}] doublet. Meanwhile, in order to show the clear linear correlation, 3206 SDSS low redshift 
($z<0.35$) quasars are selected from the database of \citet{sh11} with reliable measurements of continuum luminosity and line 
luminosities of narrow and broad H$\alpha$ and [N~{\sc ii}], and plotted in Fig.~\ref{cl}. Based on the measured emission lines 
around H$\alpha$ in \obj, the total line luminosity is $(3.04\pm0.28)\times10^{42}{\rm erg/s}$ of the emission lines around 
H$\alpha$, leading \obj~ to be an outlier in the space of continuum luminosity versus line luminosity as shown in Fig.~\ref{cl}. 
The very larger line luminosity of the measured emission lines than the expected one strongly indicate central serious obscurations. 
Second, considering the determined continuum luminosity $(1.09\pm0.31)\times10^{43}{\rm erg/s}$ at 5100\AA~ in rest frame, the 
corresponding Eddington ratio is $0.25_{-0.20}^{+1.34}$ in \obj. Then, based on the simple dependence reported in \citet{ww23} of 
Balmer decrement on accretion rate, the expected intrinsic Balmer decrement should be around 4 in \obj~ with Eddington ratio 
$\sim0.25$, smaller than the determined value 12.9. Here, the same formula is applied to determine the Eddington ratio as that 
in \citet{ww23}. Therefore, central serious obscurations should be efficiently reasonable in \obj. 

%{\color{red}Third, due to weak broad H$\beta$, through the single-epoch SDSS 
%spectroscopic properties at current stage, the \obj~ can be classified as a Type-1.8/1.9 AGN, not a standard CLAGN characterized 
%by a complete disappearance of central BLRs. The following discussed results 
%}

%%%7
	Now, lets check properties of broad Balmer emission lines of \obj~ if in bright state with weak effects of obscurations. 
If accepted E(B-V)$\sim$1.26 determined through $BD$ in the dark state, the intrinsic spectrum can be determined and shown as solid 
red line in Fig.~\ref{ebv}, after considering the following three components. The first component is from both the host galaxy and 
the narrow emission lines, with few obscuration effects. The second component and the third component are from the broad emission 
lines and from the AGN continuum emissions, with apparent and serious obscuration effects. Then, when considering obscuration effects, 
the reddening corrections are only applied to the second component and the third component. Moreover, through the reddening 
corrections with E(B-V)$\sim$1.26, the reddening corrected continuum luminosity is about $5.3\times10^{44}{\rm erg/s}$, leading 
to the expected total line luminosity to be $3.7\times10^{43}{\rm erg/s}$ of emission lines around H$\alpha$ by the relation reported 
in \citet{gh05}, which is basically consistent with the reddening corrected total line luminosity $4.5\times10^{43}{\rm erg/s}$. The 
results can be applied to further prove serious obscuration effects in \obj. Meanwhile, considering E(B-V)$\sim$1.26, the gri-band 
magnitudes are 15.68, 16.83 and 16.68, very brighter than the shown ZTF light curves in top panel of Fig.~\ref{lmc}.If in the near 
future, such bright state fortunately detected could be the direct evidence to totally support the scenario of moving dust clouds 
in \obj.

%%%8
	Furthermore, if accepted E(B-V)$\sim$0.55 in \obj, the reddening corrected spectrum is also shown as solid blue line 
in Fig.~\ref{ebv}, leading apparent magnitudes to be 17.72, 18.00 and 17.50 in ZTF gri-bands, simply consistent with the 
magnitudes of the light curves in bright state in Fig.~\ref{lmc}. Certainly, some difference between the E(B-V)$\sim$0.55 expected 
apparent magnitudes and the magnitudes in the bright state in Fig.~\ref{lmc} should be probably due to intrinsic variations of 
spectral energy distributions (SEDs) of AGN, not similar as the applied constant SEDs with considerations of different E(B-V). 
Furthermore, once accepted E(B-V)$\sim$0.55 for the ZTF light curves in bright state, the expected flux ratio is $6.8_{-2.2}^{+4.3}$ 
of broad H$\alpha$ to broad H$\beta$, leading to broad H$\beta$ and broad H$\alpha$ apparently detected in expected spectrum. 
Therefore, without considering additional effects, only considering effects of serious obscurations related to the scenario of 
moving dust clouds, type of \obj~ can be naturally changed from Type-1.8 to Type-1. In other words, the scenario of 
moving dust clouds could be reasonably applied in \obj~ as the physical origin of CLAGN. 

\begin{figure}
\centering\includegraphics[width = 9cm,height=11cm]{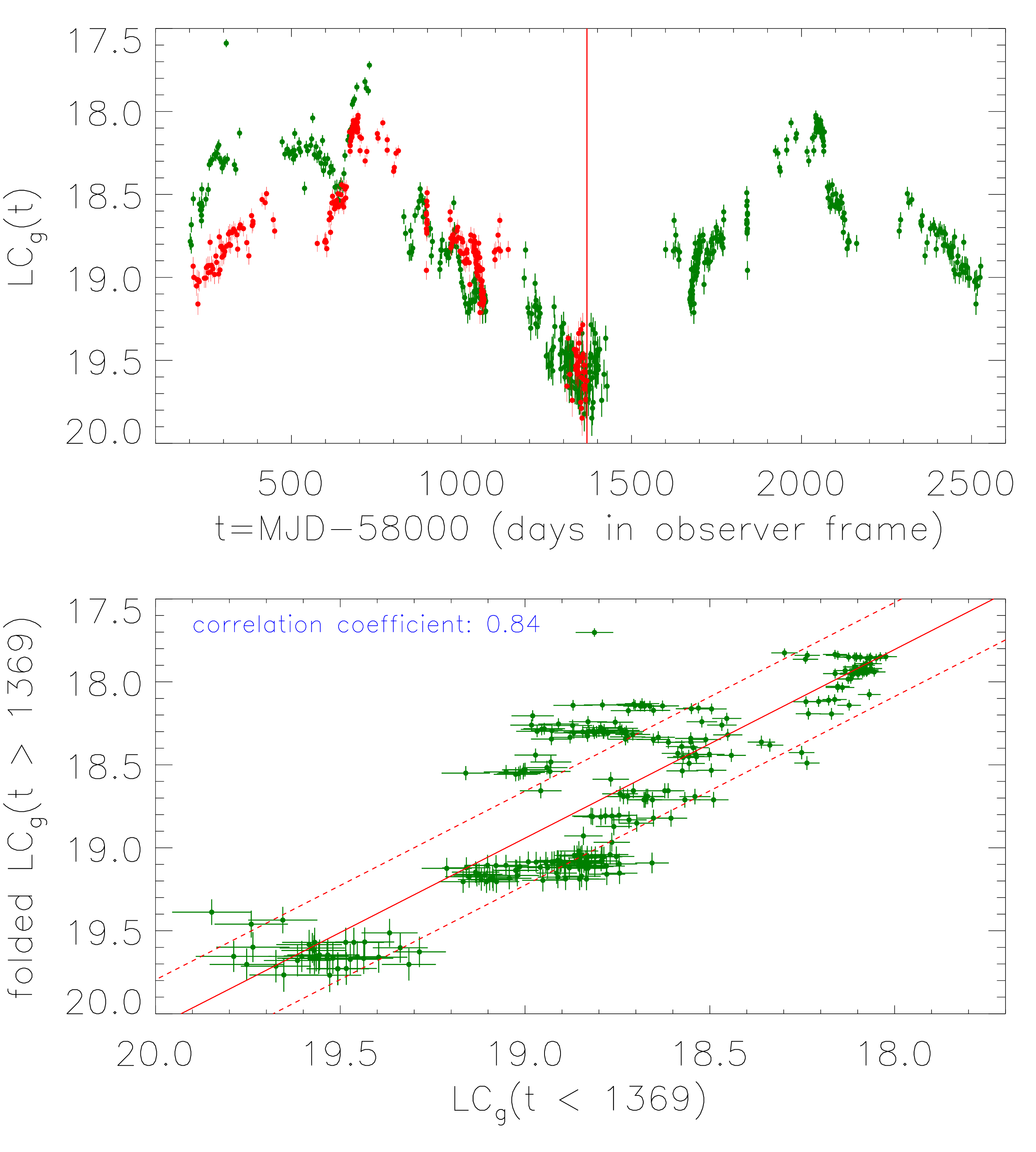}
\caption{Symmetric properties of ZTF g-band light curve in \obj. Top panel shows the ZTF g-band Light curve (symbols in dark green) 
and the folded light curve (symbols in red) with MJD-58000 larger than 1369. Vertical red line shows the symmetry axis with 
MJD-58000=1369. Bottom panel shows the correlation between the g-band light curve with MJD-58000 smaller than 1369 and the folded 
curve with MJD-58000 larger than 1369. Solid and dashed red lines show the best description and corresponding 1RMS scatters to the
correlation by Y=A+B*X.}
\label{a1}
\end{figure}

\begin{figure}
\centering\includegraphics[width = 9cm,height=11cm]{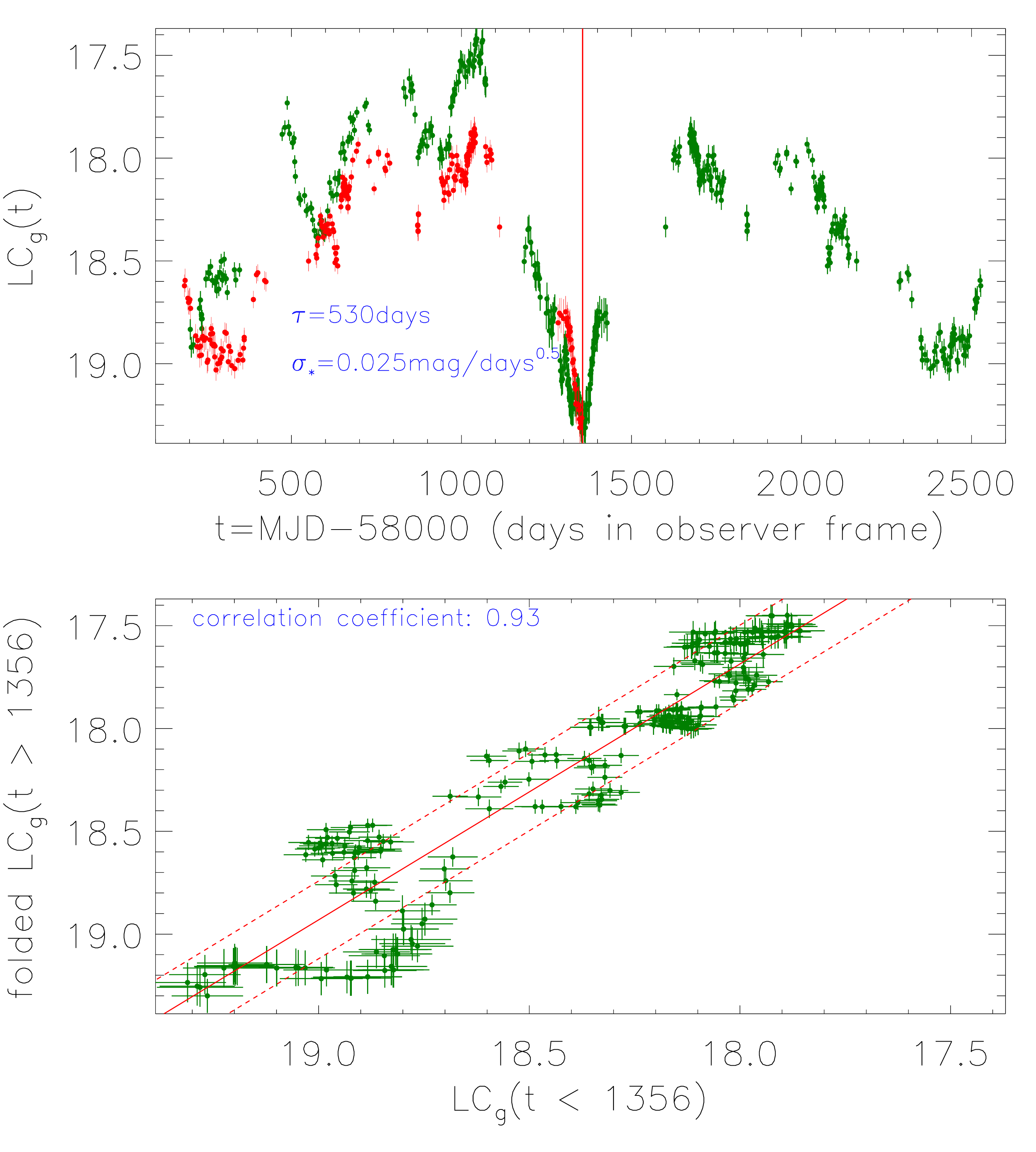}
\caption{Symmetric properties of one fake light curve created by the CAR process. Symbols and line styles have the same meanings 
as those in Fig.~\ref{a1}. In top panel, the applied CAR process parameters of $\tau$ and $\sigma_*$ have been marked in blue 
characters for the shown fake light curve.}
\label{a2}
\end{figure}

%%%9
	Before ending the section, seven additional points are noted. First, at current stage, it is hard to rule out the scenario 
of variations in accretion rates in \obj, especially due to large scatters of the dependence reported in \citet{ww23} of Balmer 
decrement on accretion rate. Variability of broad emission lines in multi-epoch spectra should provide further clues to support or 
to be against the preferred scenario of moving dust clouds in \obj. If the variations of accretion rates were preferred, the trend 
could be expected that weak continuum luminosity leads to wider broad emission lines, due to virialization assumption efficiently 
applied in central BLRs \citep{pe04}. However, if accepted the scenario of moving dust clouds, there should be no such trend or only 
a very weak trend. To check variability of broad emission lines in \obj~ is our main objective in the near future.

%%10
	Second, although the scenario of moving dust clouds is preferred in \obj, it is not clear for the origin or the spatial 
structures of the clouds. However, after checking the light curves, it can be confirmed that the light curves in Fig.~\ref{lmc} are 
symmetric about the vertical line MJD-58000=1369 (time of the darkest point). As shown in Fig.~\ref{a1}, top panel shows the results 
on the symmetric g-band light curve of \obj~with MJD-58000=1369 as the symmetry axis, bottom panel shows the correlation with 
correlation coefficient 0.84 ($P_{null}<10^{-10}$) between the g-band light curve with MJD-58000 smaller than 1369 and the folded 
curve with MJD-58000 larger than 1369. Therefore, symmetric spatial structures can be expected for the clouds in \obj, consistent 
with expected results by the symmetric Gaussian-like descriptions to the colors in bottom panel of Fig.~\ref{lmc}. 

%%%11
	Third, as discussed in \citet{lc15, run16}, for foreground moving dust clouds in a bound orbit around the central BH in 
front of the continuum emission source and BLRs, the crossing time for the moving dust clouds can be estimated by 
\begin{equation}
t_{cross}\sim0.07R_{orb}^{1.5}M_{8}^{-0.5}\arcsin(R_{src}/R_{orb})
\end{equation}.
In the equation above, $R_{orb}$, $M_8$, $R_{src}$ are the distance between the foreground moving dust clouds and the central BH, 
the BH mass in units of $10^{8}{\rm M_\odot}$ and the true lateral scale size of the BLRs. It is necessary to to check whether 
the timescale 3-4 years for the variations related to moving dust clouds in \obj~ is long enough relative to the $t_{cross}$. If assuming 
$R_{orb}\sim k_o\times R_{BLRs} (k_o>1)$ and $R_{src}\sim k_s\times R_{BLRs}$ with $R_{BLRs}$ as the distance between central BLRs 
and central BH, the cross timescale can be estimated as
\begin{equation}
	t_{cross}\sim2.9M_8^{-0.5}k_o^{1.5}\arcsin(k_s/k_o){\rm years}
\end{equation},
after accepted the intrinsic $R_{BLRs}\sim12$ light-days through the reddening corrected (E(B-V)$\sim$0.55) intrinsic continuum  
luminosity $1.4\times10^{43}{\rm erg/s}$ at 5100\AA~ in rest frame applied in the empirical relation in \citet{ben13}. If simply 
accepted that $k_o\sim1.5$ (moving dust clouds lying outside of the central BLRs) and $k_s\sim1/4$ (central BLRs having not large 
lateral scale size), we will have $t_{cross}\sim3.8$ years, consistent with the variability timescale 3-4 years in \obj. In other 
words, the results on crossing timescale cannot provide apparent clues to be against the scenario of moving dust clouds in \obj.

%%%12
	Fourth, we give a simple way to roughly estimate how rare this kind of objects like \obj~ appears in SDSS. Based on the 
selected 3206 SDSS low redshift quasars from the database of \citet{sh11} as shown in Fig.~\ref{cl}, there are only 1 quasar, the 
\obj, having its locations lying above the 5RMS scatters of the strong linear correlation. Therefore, only through spectroscopic 
properties, the expect detection rate is about 0.03\% (1/3206) for objects similar as the \obj, to support that the \obj~ is unique 
enough. Here, we should note that there are 5 objects having their locations lying below the 5RMS scatters of the strong linear 
correlation, as shown in Fig.~\ref{cl}. However, the reddening corrections cannot be applied to explain expected line luminosities 
very larger than the measured ones. Therefore, there are no further discussions on the 5 objects.

%%%13
	Fifth, there are no considerations on different extinction between the BLRs and the continuum emission sources. If accepted 
different extinctions, there should be more heavily reddening effects on continuum emission sources than on central BLRs, leading to 
one object having more pronounced deviation from the strong linear correlation shown in Fig.~\ref{cl}. In other words, considering 
different extinctions should be more conducive to finding objects having similar properties as the \obj~ in the space of continuum 
luminosity and line luminosity. Meanwhile, the different extinctions should indicate different E(B-V) applied on continuum emissions 
from the E(B-V) applied on broad line emissions, leading to different continuum emission intensities from those shown in Fig.~\ref{ebv} 
after considering type changes related to reddening corrections. However, the scenario of moving dust clouds can also be efficiently 
applied for the variations in broad emission lines.

%%%14
	Sixth, based on the symmetric properties of light curves, a simple method can be applied to check the probability of such 
symmetric light curves related to variations of intrinsic accretion rates in \obj. Similar as what we have recently done in 
\citet{zh25c}, the continuous autoregressive (CAR) process is applied to describe light curves related to intrinsic variations 
of accretion rates as discussed in \citet{kbs09, koz10, mac10}. Based on the CAR process, light curves $LC(t)$ related to AGN 
variability can be artificially simulated by
\begin{equation}
dLC(t)~=~\frac{-1}{\tau}LC(t)dt~+~\sigma_*\sqrt{dt}\epsilon(t)~+~LC_0
\end{equation}.
In the equation above, $\tau$ (in units of days) and $\sigma$ (in units of ${\rm mag/days^{0.5}}$) are the process parameters, 
$LC_0$ is the mean value of $LC(t)$, and $\epsilon(t)$ is a white noise process with zero mean and variance equal to 1. Here, we have 
accepted the time information (MJD-58000) and the mean magnitude of the ZTF g-band light curve of \obj~ as the $t$ and $LC_0$ in 
the equation above. Then, through the randomly selected values of $\tau$ from 100 to 1200 and $\sigma_*$ from 0.006 to 0.03 (common 
values of $\tau$ and $\sigma_*$ in quasars), there are $10^5$ fake light curves to be created to trace variability related to 
variations in accretion rates. Then, the following three criteria are applied to select fake light curves having the similar 
symmetric properties as those of \obj. First, the correlation coefficient should be larger than 0.84 (the value for the g-band 
light curve of \obj) between the fake light curve with MJD-58000 smaller than the time of the symmetric axis (time of the darkest 
point in the fake light curve) and the folded curve with MJD-58000 larger than the time of the symmetric axis. Second, the amplitude 
of the fake light curve should be larger than 1.3mags (the value for the g-band light curve of \obj). Third, the time (MJD-58000) 
of the symmetric axis should be larger than 1200 and smaller than 1700, in order to avoid the cases with too small number of data 
points on one side of the symmetry axis. According to the three criteria above, among the $10^5$ fake light curves, there are 3166 
fake light curves selected to have the similar symmetric properties as those of the ZTF g-band light curve of \obj. Meanwhile, 
Fig.~\ref{a2} shows an example on the fake light curve created by the CAR process, leading to symmetric variability properties. 
Therefore, the probability is smaller than 3.2\% (3166/$10^5$) that the symmetric properties in the light curves of \obj~ are related 
to variations of accretion rates, which will be accepted indirect evidence to support the scenario of moving dust clouds.

\section{Conclusions}

	Motivated by the unique photometric variability properties of ZTF gri-band light curves of \obj~ with large variations in 
optical colors to support central obscurations with confidence level higher than 10$\sigma$, the scenario of moving dust clouds is 
proposed to explain probably expected changing-look properties of \obj. Based on the measured broad emission lines of \obj, the 
serious obscured spectroscopic results are consistent with expected results from the photometric light curves at the dark state. 
Meanwhile, based on the ZTF gri-band light curves from dark state to bright state, the expected flux ratio of broad H$\alpha$ to 
broad H$\beta$ should be changed from about 12.9 to 6.8, leading \obj~ to be a normal type-1 AGN, after accepted reddening corrections. 
Furthermore, after checking the symmetric variability properties of photometric light curves if related to variations of intrinsic 
accretion rate, the probability is 96.8\% (1-3.2\%) to support the scenario of moving dust clouds preferred in \obj. Therefore, 
probably in a sample of CLAGN with serious obscurations expected by optical colors, the scenario of moving dust clouds should be 
the main physical mechanism. It will be helpful for moving dust clouds efficiently applied in some CLAGN by detecting more AGN 
having optical variability properties similar as those in \obj~ in the near future.

\section*{acknowledgements}
Zhang gratefully acknowledge the anonymous referee for giving us constructive comments and suggestions to greatly 
improve the paper. Zhang gratefully thanks the kind financial support from HangJi Action Plan under the Guangxi Science and 
Technology Program 2026GXNSFDA00640018 and from NSFC-12373014 and NSFC-12173020. This manuscript has made use of the data from 
the SDSS (\url{https://www.sdss.org/}) and ZTF(\url{https://www.ztf.caltech.edu/}).

%\section*{Data Availability}
%The data underlying this article will be shared on reasonable request to the corresponding author
%(\href{mailto:xgzhang@njnu.edu.cn}{xgzhang@njnu.edu.cn}).

\end{document}